\documentclass[global,twocolumn,referee]{svjour}
\usepackage{graphics}
\journalname{Applied Physics A}
\begin{document}
\title{Bandstructure and optical properties of $\alpha-LiIO_{3}$ crystal}
\author{Yung-mau Nie}
\thanks{\emph{Present address: Computational Materials Science Center, National Institute for Materials Science, Sengen 1-2-1 Tsukuba, Ibaraki, Japan, 305-0047, Tel/Fax:+81-29-858-8003}}
\institute{Institute of Physics, Academia Sinica, 128 Sec. 2,
Academia Rd, Nankang, Taipei 115, Taiwan, R.O.C.,
\email{ymnie@phys.sinica.edu.tw}}
\date{Received: 26 February 2007/ Revised version: date}
\maketitle
\begin{abstract}
The bandstructure was calculated by the full-potential linearized
augmented plane wave method. The result reveals two important
insights to the novel second harmonic generation (SHG) of
alpha-phase lithium iodate ($\alpha-LiIO_{3}$) crystal: the
existence of finite intra-band momentum matrix elements due to the
non-inversion symmetry of the crystal illuminating the potential of
the intra-band transition, and the strong covalent bonding between
the $I$-atoms and the ligand $O$-atoms resulting the condition of
the double-resonance. An inter-band transition scenario in SHG as
$\alpha-LiIO_{3}$ in nano-structure is proposed. The optical
properties were calculated within the theoretical framework of the
time-dependent perturbation of the independent-particle model. The
dielectric tensors and the refractive index were evaluated.
Comparisons between the predictions and the results were made: the
x-ray near edge absorption spectra; the refractive index at the
static limit, and at finite frequencies. Possible factors
attributing the calculation errors is discussed. 
\end{abstract}
\section{\label{sec1}Introduction}
\begin{figure}[htbp]
\resizebox{0.45\textwidth}{!}{%
  \includegraphics{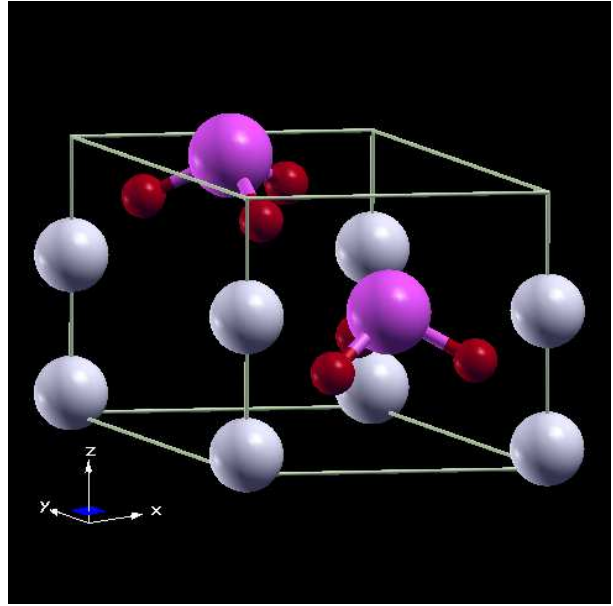}}
\caption{\label{fig:struct}The unit cell structure of the
$\alpha-LiIO_{3}$ crystal. In the bonding sketch, each $I$-atom
(purple color) locates at the top of pyramid based by three
$O$-atoms (red color). The $Li$-atoms (grey color)reside the sites
on the edge.}
\end{figure}

\indent The alpha-phase lithium iodate ($\alpha-LiIO_{3}$), in a hexagonal crystal as depicted in Fig.~\ref{fig:struct} \cite{Svensson,Rosenzweig}, has been intensively studied in the past years due to its novel nonlinear optical \cite{Singh}, piezoelectric \cite{Haussuhl,Warner}, and acousto-optic \cite{Warner,Aliev,Vorobev} properties. Its nonlinear optical property, especially in the second harmonic generation (SHG), always receives the attentions from the electro-optical technology, which is recently rising up due to the all-optical information processing for modern telecommunications systems. Furthermore, very recently the development of $\alpha-LiIO_{3}$-based nano-structural systems for the nonlinear optical waveguide is very promising because of the preservation for the novel nonlinear optical functions in the bulk. \cite{Teyssier1,Teyssier2} In addition, the developments of photorefractive devices in real-time holography, and image processing, also urgently demand the ultraviolet photorefractive property of $\alpha-LiIO_{3}$ to improve resolution and storage capacity. \cite{Gunter} With the analogous characteristics: high refractive index, excellent optical transmittance in the visible and near-infrared region, and high dielectric constant, the newly invented room-temperature ferromagnetic semiconductor, $TiO_{2}$ in anatase structure under a $Co$-doping \cite{Matsumoto}, has inspired the spintronic technology toward the magneto-optic and the optoelectronic applications. Thus presumably, the $\alpha-LiIO_{3}$ crystal should have the potential to be developed as magnetically doped reformations or a associated substrate material in the same way. In spite of this material being invented very early, the theoretical study at First-principles level on the electronic structure and the optical transition is surprisingly very limited. The resulting characters of bandstructure can further guide the searching and the tailoring of new non-linear optical materials.\\
\indent The present First-principles calculations were performed by
the full-potential linearized augmented plane wave (FLAPW) method of
density functional theory defined by the local density approximation
(DFT-LDA). \cite{Andersen,Blaha1,Blaha2,Blaha3,Madsen,Cottenier}
Testing by the f-sum rule \cite{Altarelli,Smith}, the eigen-values
and the momentum-matrix elements was verified to be valid to produce
correct evaluations of optical properties. By virtue of the analysis
of the band-structure, the insights for the intra-band and
inter-band transitions of the remarkable SHG property are
discovered. The near-edge states dominating the optical transition
were further analyzed. The conclusion of it supplies an useful
scenario in the classical dipole-oscillation framework to understand
the dielectric response of the $\alpha-LiIO_{3}$ crystal.
In order to comparing with the experimental results about the bandstructure, a simulation of the x-ray near edge absorption spectra was incorporated here.\\
\indent On the calculation of optical functions, the time-dependent
perturbation of independent particle model at the long-wavelength
limit was applied, in which the eigen states of quasi-particle were
approximated as those given by the DFT-LDA. In addition, in a
non-perturbative way the time-dependent DFT , allowing for the
ab-initio calculations on electron in external electromagnetic
fields, also solves the analogous problem, so it is able to capture
the strong dynamic effect of system illuminated in the high power.
\cite{Marques,Corso} On the other hand, the recently well-developed
Berry phase method \cite{Vanderbilt} can also directly simulate the
macroscopic polarization at the ab-initio level to the external
field effect. Due to a greater demand of the computation, so far
they are more specified to the nonlinear optical simulations
involved significant dynamic effect and the polarization in
ferroelectric systems, respectively. However, the relation between
the bandstructure information and the optical transition mechanisms,
directly inspiring the tailoring the electronic structure of
materials, can be feasibly accessed by the perturbation method.
Hence, plenty previous works applied it to calculate the dielectric
and SHG tensors of transparent
insulators.\cite{Sipe0,Sipe1,Sipe2,Sipe3,Sipe4,Sipe5,Rashkeev,Sharma,Adolph,Duan1,Duan2}
So far comparing with experimental results, such a method makes good
agreement for the dielectric response calculations.\\
\indent The present article is outlined as follows. In
Sec.~\ref{sec2}, the implementation of the FLAPW method and the
formalism for the dielectric response will be illustrated. In
Sec.~\ref{sec3}, the resulting bandstructure and the dielectric
functions will be exhibited. The insights for the novel SHG property
and the optical transition  structure will be discussed. In
Sec.~\ref{sec4}, the comparisons with the experiments on the x-ray
near edge absorption spectra, the refractive index at the static
limit and the finite frequency will be given therein. Finally, a
brief summary will be given in Sec.~\ref{sec5}.
\section{\label{sec2} METHODOLOGY AND FORMALISM}
\subsection{FLAPW Method}
\indent
The modified FLAPW method, 'APW+lo' \cite{Madsen}, was
applied via the implementation of the WIEN2K code. \cite{Blaha1,Blaha2,Blaha3,Madsen} The exchange-correlation potential
functional was defined by the generalized gradient approximation
(GGA) parameterized by Perdew and Wang \cite{Perdew}. The core and
the valence states were respectively calculated relativistically and
semi-relativistically. The muffin-tin radii were set to be $1.8$,
$1.83$, $1.6$ {\AA} for the $Li$, the $I$, and the $O$-atom,
respectively. The expansions of associated Legendre polynomials for
spherical harmonics of the wave function and of the non-spherical
full-potential expansion were truncated at $l=10$ and $l=4$,
respectively. The parameter $RK_{max}$ was set to be $8.5$.
Additional local orbitals were added to incorporate low-lying
valence states in the semi-core regime: $4d$-states of the
$I$-atoms, $2s$-states of the $Li$ and of the $O$-atoms. The lattice constants $a$ and $c$ at energy minimum, $5.574$ and $5.259$, were obtained by the parabolic fitting of bulk-modulus calculation, and the atomic coordinates: $Li(0.0, 0.0, 0.07145)$; $I(0.3333, 0.6666, 0.9991)$, and $O(0.2470, 0.3422, 0.8379)$ at the equilibrium were determined by the calculation of ionic relaxation. Those resulting lattice parameters for the present bandstructure calculation are very similar to the x-ray diffraction \cite{Svensson} reported previously.
\subsection{FORMALISM}
\indent
In the independent particle model, the optical conductivity function $\sigma$ is expressed as \cite{Li}
\begin{eqnarray}
\sigma_{aa}(\omega) &=& \frac{2\pi}{\omega \Omega}\int \frac{d \textbf{ k}}{4\pi^{3}}\sum_{n,m}|p_{nm}^{a}|^{2} \delta (\omega_{n}-\omega_{m}-\omega); \label{eq:optc}\\
\delta(x) &\equiv &\frac{1}{\sqrt{\pi}\Gamma}e^{-(\frac{x}{\Gamma})^{2}},
\end{eqnarray}
where $\omega_{n}$ is the energy of the $n$-th band; $\Omega$ denotes the volume of unit cell; $\omega$ is the photon energy, and $p_{nm}^{a}$ is the $a$-component of the momentum matrix element in the Cartesian coordinate.
Herein, the subscript $n$ ($m$) symbolizes a conduction (valance) band.
The $\delta$ function was defined as a smooth Gaussian distribution with the $\Gamma$ of $0.35$ eV.
Thus, the imaginary part of dielectric tensor $\varepsilon_{2}$ can be obtained by the relation: $\varepsilon_{2}(\omega)=(2\pi/\omega)\sigma(\omega)$.
By means of Kramers-Kronig transform, the real part $\varepsilon_{1}$ and the refractive index $n(\omega)$ can be obtained as follows
\begin{eqnarray}
\varepsilon_{1}(\omega)&=&1+8 P \int_{0}^{\infty} \frac{\varepsilon_{2}(\omega^{'})}{{\omega^{'}}^{2}-\omega^{2}} d\omega^{'}, \label{eq:K-K}\\
n(\omega)&=&(\frac{\sqrt{\varepsilon_{1}^{2}(\omega)+\varepsilon_{2}^{2}(\omega)}+\varepsilon_{1}(\omega)}{2})^{1/2}. \label{eq:refractive}
\end{eqnarray}
\indent
The Brillouin-zone integration are achieved by the irreducible points of special-point sampling \cite{Monkhorst}.
The convergence of the integration was tested. The error range was estimated to be at least less $0.1$ percent by obtaining difference between two results given by the $12 \times 12 \times 11$ and the $23 \times 23 \times 21$ mesh, respectively.
\section{\label{sec3} RESULTS AND DISCUSSIONS}
\subsection{BANDSTRUCTURE}
\begin{figure}[htbp]
\resizebox{0.5\textwidth}{!}{%
  \includegraphics{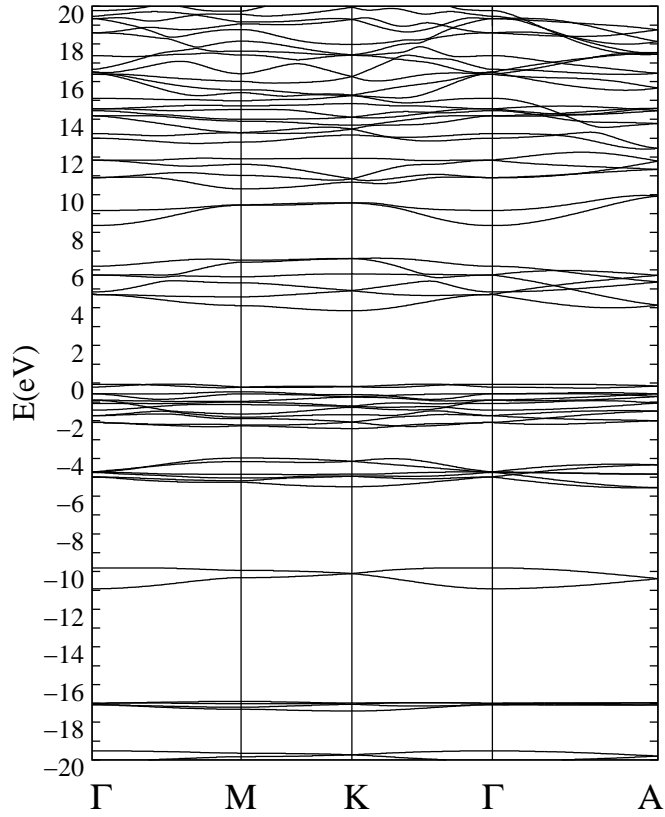}
} \caption{\label{fig:bandst} The bandstructure of the
$\alpha-LiIO_{3}$ crystal. Here $\Gamma: (0,0,0)$; $M:
2\pi/a(1/2,0,0)$; $K: 2\pi/a(1/3,1/3,0)$, and $A:2\pi/c(0,0,1/2)$.}
\end{figure}

\begin{figure}[htbp]
\resizebox{0.45\textwidth}{!}{%
  \includegraphics{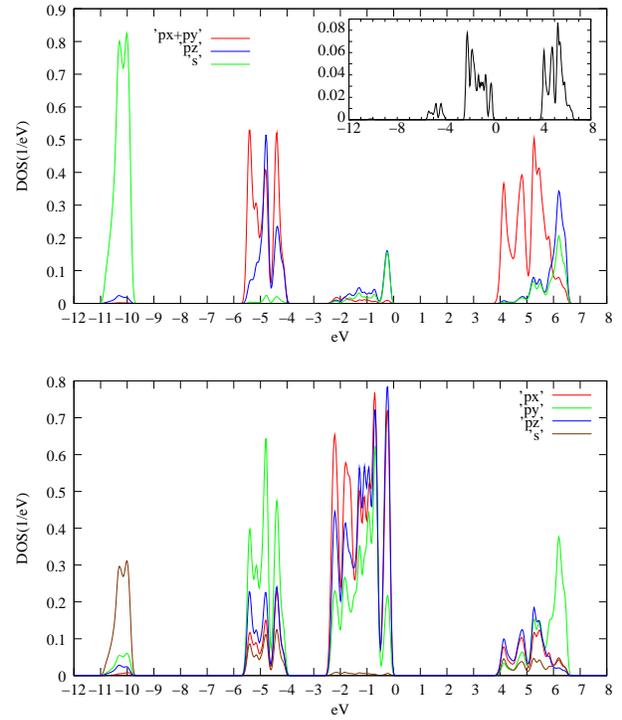}
} \caption{\label{fig:PDOS} The LDOS for the s- and p-states of the
$I$-atom (top) and the $O$-atom (bottom). The inset of the top panel
depicts the result of d-states of $I$-atom.}
\end{figure}

\begin{figure}[htbp]
\resizebox{0.4\textwidth}{!}{%
  \includegraphics{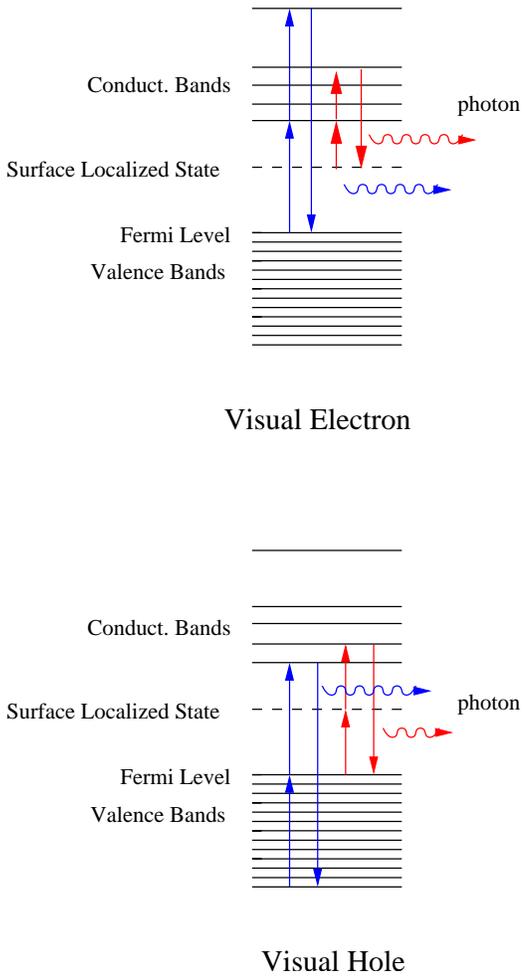}
} \caption{\label{fig:SHG} The visual hole and the visual electron
mechanims of the inter-band transition in SHG. The blue arrow
depicts the original transition in the bulk, and the red specifies
to the case for the bands of surface localized states only.}
\end{figure}

\indent
Firstly, the remarkable feature, all resulting bands within the range $-6$ to $0$ eV always appear in a number of pairs slightly split from one degenerate state, is distinguishable in the Fig.~\ref{fig:bandst}. The resulting local density of state (LDOS), depicted in Fig.~\ref{fig:PDOS}, indicates them to be derived by the $O-O$ and $I-O$ bonding. With respect to this point, according to the '$\textbf{ k} \cdot p$' method, the band dispersion about the $\Gamma$-point can be approximated to the second order in $\Delta \textbf{ k}$ as\cite{Lax}
\begin{eqnarray}
E_{m}(\textbf{ k}+\Delta \textbf{ k})&=&E_{m}(\textbf{ k})+\frac{\hbar}{m} \Delta \textbf{ k} \cdot p_{mm}\nonumber+\frac{\hbar^{2}(\Delta \textbf{ k})^{2}}{2m}\\
&+&\frac{\hbar^{2}}{m^{2}}\sum_{n \neq m}\frac{(\Delta \textbf{ k} \cdot p_{mn})(\Delta \textbf{ k} \cdot p_{nm})}{E_{m}(\textbf{ k})-E_{n}(\textbf{ k})}.
\end{eqnarray}
In a state with inversion symmetry, the parity of any its physical
expectation is even, so the intra-band momentum matrix element
$p_{mm}$ vanishes due to the odd parity of the momentum operator.
Contrarily, in a state without the inversion symmetry, a finite
$p_{mm}$ exists for the opposite situation. The former causes the
first order term in the expansion to vanish so to give a perfect
parabolic band curvature; however, the later results a non-zero
linear dependent energy-split added to the parabolic band. The scale
of $p_{mm}$ should be much less than the $\hbar \Delta \textbf{ k}$,
deduced from the resulting energy-split appearing to be very narrow. Then such a fine
structure character can be viewed as the finger-print of the
structural non-inversion symmetry of the $\alpha-LiIO_{3}$ crystal.
In fact, about the SHG the existence of finite $p_{mm}$ subjects the intra-band transitions \cite{Sipe5,Rashkeev}, though prohibited in the optical transition, so the present work reveals the potential of this transition mechanism to generate certain contribution.\\
\indent
Secondly, there is a large energy-split around $10$eV resulted by the covalent bonding between the $I$-atoms and the ligand O-atoms according to the resulting LDOS, implies an extremely strong interaction associated with the bond forming. Actually it is even greater than the magnitude of the strong on-site Coulomb repulsion $U$ in most of transition-metal or rare-earth atoms in the perovskite crystals \cite{Solovyev,Anisimov}, also widely applied as nonlinear optical materials \cite{Singh}. Furthermore, the derived double-gap feature exhibited in the Fig.~\ref{fig:bandst}, separating the $I-O$ bonding and anti-bonding states, as well as the intervening states localized on the $O$-atom near the Fermi-surface, naturally meets the isometric inter-band spacing condition for the double-resonance of the inter-band transition in SHG. However, the $Li$-atom was determined to make only little contribution to the aforementioned states.\\
\indent
Predictably in the nano-structure system some surplus bands were induced within the gap because of the incorporation of surface localized states. According to the double-resonance scenario of the visual hole or the visual electron mechanism \cite{Sipe0} illustrated in Fig.~\ref{fig:SHG}, the original novel SHG in the bulk can be still preserved; however, those bands of surface localized states, unless they near the gap-edge, would be hard to give a significant impact. On the other hand, after all presumably the number of the bands from surface localized states is much less than that from near-edge valence states in the bulk, also make it not be a major role in SHG. Such concepts should be useful to figure out the novel nonlinear optical properties in the bulk still preserving in the nano-structured systems. \cite{Teyssier1,Teyssier2}\\
\indent The present calculation indicates the band gap as the type
of allowed indirect transition. Based on the resulting LDOS, the
permitted transition between the gap-edge bands indeed occurs within
each atomic muffin-tin sphere, according to the angular momentum and
parity selection rules of the atomic spectroscopy.
The obtained indirect type agrees with the previous experimental conclusion \cite{Gaffar}. It is deduced for the sake of the directionality of the near-edge p-states tending to maximize the band dispersion at the $\Gamma$-point, and to minimize it at the zone corner.\cite{Singh-book} The obtained gap value, $3.8$ eV, is slightly less than the measured edge-onset, $\simeq 4$eV, in the absorption \cite{Regel,Xu,Gaffar,Galez} and the transmission spectrum \cite{Nash}.\\
\indent
The test of the f-sum rule \cite{Altarelli,Smith}, $\int_{0}^{\infty}\frac{2\Omega}{\pi}\sigma(\omega)d\omega=\Sigma_{i}f_{i}=N_{eff}$, was performed to exam the obtained bandstructure and the momentum matrix elements. For the very less contribution, the $d$-electrons of the $I$-atoms should be excluded, so the effective valence electrons is $52$, being compatible with the results of $N_{eff}$: $56.83$ and $56.66$ for the $xx$ and the $zz$-component, respectively. This implies the obtained bandstructure and momentum matrix elements to be amenable to the following optical calculations.\\
\subsection{DIELECTRIC RESPONSE}
\begin{figure}[htbp]
\resizebox{0.45\textwidth}{!}{%
  \includegraphics{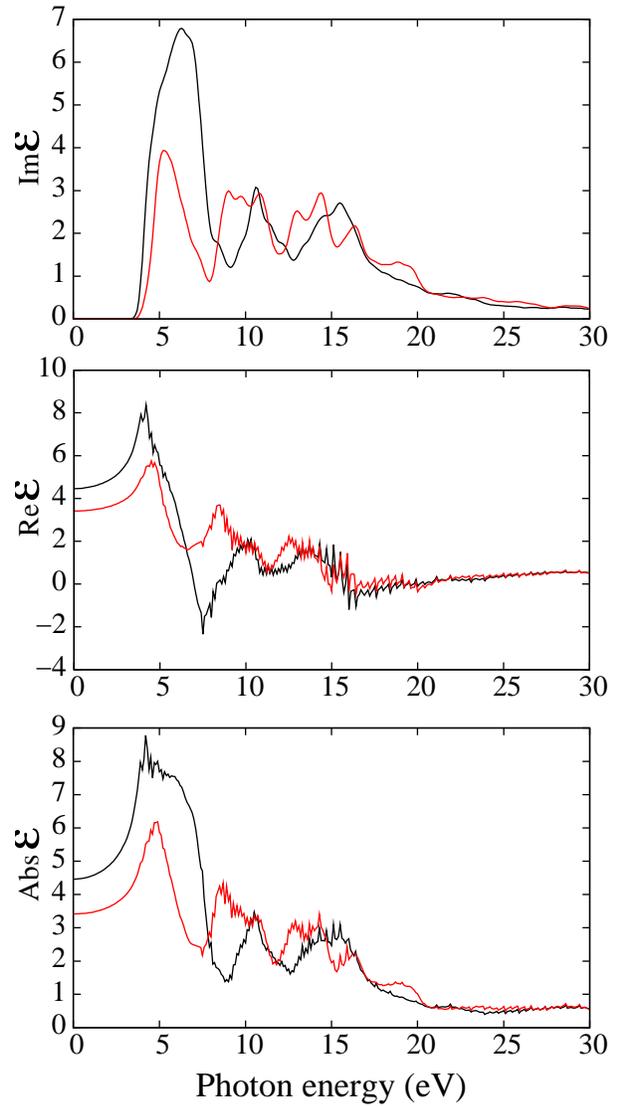}
} \caption{\label{fig:eps} The dispersion of the imaginary part, the
real part, and the absolution of the dielectric tensor (in the unit
of $(eV \cdot sec)^{-1}$). The black and the red lines depict the
$\varepsilon_{xx}$- and the $\varepsilon_{zz}$-components of the
dielectric function, respectively.}
\end{figure}

\indent The resulting components of the dielectric tensor are
exhibited in Fig.~\ref{fig:eps}. In fact, the resulting
$yy$-component is identity to the $xx$-one, consisting with the
experimental observations. The significant optical anisotropy
behaves as the obviously different dispersions between the $xx$- and
the $zz$-component. It is actually dominated by the discrepancy of
respective strength component, $|p_{nm}^{a}|^{2}$, in the
Eqn.(~\ref{eq:optc}). Though the information of bandstrucure is hard
to directly access the insight of this quantity, it still supplies
the optical transition knowledge: the initial state almost
localizing on the $O$-atoms and the final state mainly derived from
the $I-O$ anti-bonding states. In fact, the absorption
resonance-edge of the resulting imaginary part given by the gap
separating the above two classes states, and the location of the
absorption peak, $7$eV, equivalent to the energy-split between them,
further solidate the aforementioned state information in the
transition. Taking advantage of the classically electromagnetic
radiation concept, the dynamical distribution of the charge-density
in transition like a dipole rapidly oscillating out of phase along
the $I-O$ bond, is useful to figure out the resulting optical
anisotropy. It indeed gives an isotropic radiation on the $xy$-plane
as the calculated result.
\section{\label{sec4} COMPARISON WITH EXPERIMENTS}
\begin{figure}[htbp]
\resizebox{0.45\textwidth}{!}{%
  \includegraphics{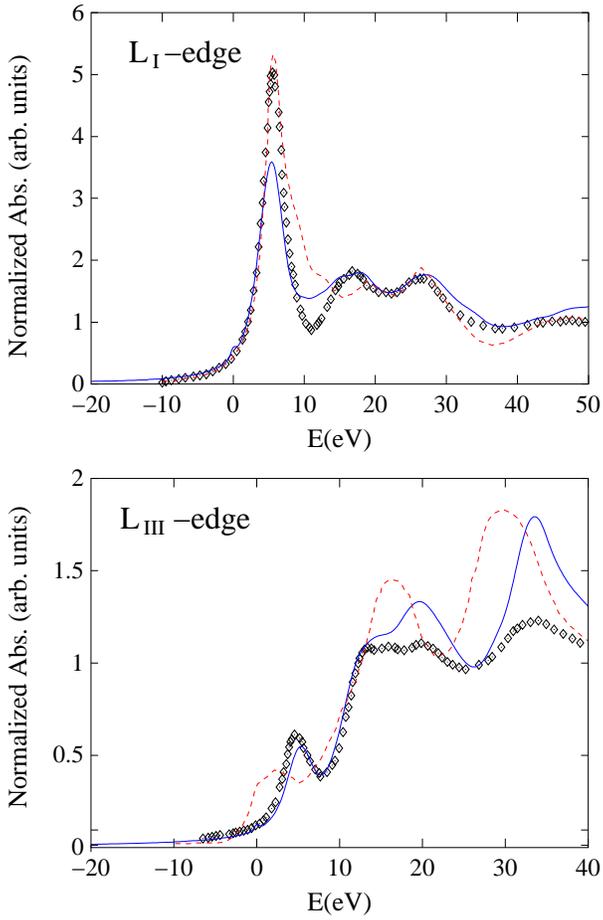}
} \caption{\label{fig:XANES} The X-ray absorption near edge spectra
of the $I$-atom at the $L_{I}$, and the $L_{III}$-edge. The present
results, and the corresponding measurements; their associated
simulations of the multiple scattering theory are respectively
depicted by the blue solid line, and the diamond symbols; the red
dashed line.}
\end{figure}

\begin{figure}[htbp]
\resizebox{0.45\textwidth}{!}{%
  \includegraphics{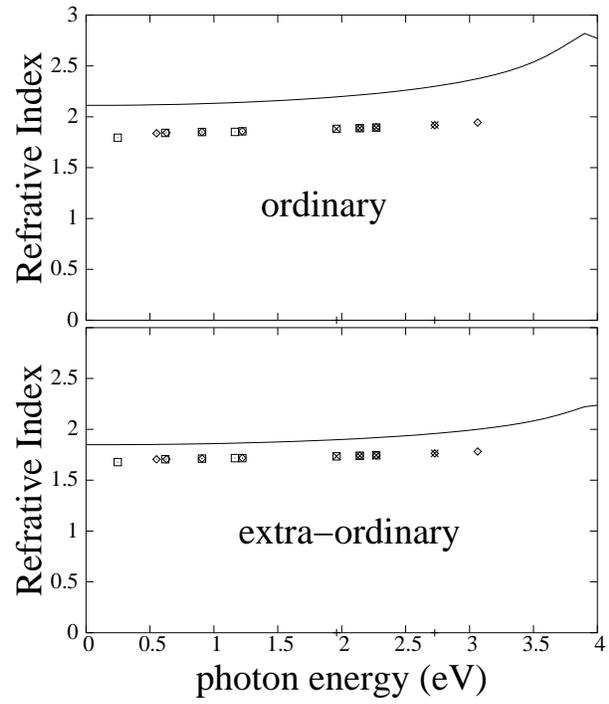}
} \caption{\label{fig:refra} The dispersion of the refractive index.
The solid line depicts the present result and the diamond, the
cross, the square, and the triangle symbols exhibit the measurements
of references \cite{Umegaki}, \cite{Crettez}, \cite{Herbst}, and
\cite{Takizawa}, respectively.}
\end{figure}

\indent
The resulting X-ray absorption near edge spectra of the $I$-atom at the $L_{I}$, and the $L_{III}$-edge are shown in Fig.~\ref{fig:XANES}. The used parameters of resolution identity to experimental values: $0.75$, and $0.66$ eV \cite{Goulon}; the inserted values of atomic natural widths are $3.46$, and $3.08$ eV \cite{Krause} for the $L_{I}$- and the $L_{III}$-edge, respectively.
The present calculation agrees well with the previous measurements and their associated calculations of the multiple scattering theory \cite{Goulon}, especially in the near edge regime.
Such as the experimentally observed white line feature at the $L_{I}$-edge, as well as the previously discovered pre-edge structure at the $L_{III}$-edge were reproduced.
However, the discrepancy at the $L_{I}$-edge might be due to the resulting $5${\it p}-hybridization to be under-estimated.\\
\indent The resulting refractive indices and the corresponding
experimental results are listed in Table ~ \ref{tab:refra}. In fact,
the adopted formula all are approximated at the long-wavelength
limit, so the most adequate comparison to the experimental data, to
drastically rule out the local field and the dynamic effects, should
be right at the static limit. The constant, measured from a number
of independent experiments \cite{Nath,Umegaki,Takizawa}, is
reproducible in the current calculation. Besides, the remarkable
experimental negative birefringence of this crystal is also
reproduced here.
Since a slightly over-estimated ordinary and extra-ordinary component was produced here, this causes the disagreement with experimental values in the negative birefringence. Previous publications suggested the causes of discrepancy originating from the defect in details of the bandstructure \cite{Lambrecht,Levin43}, so the unincorporated non-local effect of the exact density functional \cite{Aulbur} is deduced to make certain influence to them.\\
\indent
The resulting dispersion of the ordinary $n_{o}$ and the extra-ordinary refractive index $n_{e}$ at finite frequencies are shown in the Fig.~\ref{fig:refra}.
Generally, the present results perfectly match the frequency-dependent trend and give a consistent deviation with respect to the experimental measurements in the transparent regime.
Some degree of discrepancy from the experiments might be the result of un-incorporations of the aforementioned non-local effect and the dynamic many-body effect, and the local field effect.\\
\section{\label{sec5}SUMMARY}
The truth of the bandstructure and the validity of the momentum
matrix elements given by First-principles calculations are
respectively verified by the comparison with the previous
experimental result of X-ray absorption near edge spectra, and the
test of the f-sum rule. The existence of finite intra-band momentum
matrix elements due to the non-inverse symmetry of the crystal is
revealed by the resulting bandstructure, which dominates the
intra-band transition in SHG. In addition, the energy-split of the
$I-O$ bonding results the condition of the double-resonance of the
inter-band transition in SHG. The suggested scenario of the
double-resonance in SHG as $\alpha-LiIO_{3}$ in the nano-structure
is an useful reference to the tailoring of a same type of
nonlinear optical systems. The present simulation, basing on the
time-dependent perturbation in independent-particle model, can well
capture the features of the linear dielectric response of the
$\alpha-LiIO_{3}$ crystal. The discrepancies from the comparison
with experiments are deduced from the ignored non-local effect in
calculating the bandstructure, and the unincorporated many-body
dynamic and local field effects in evaluating the dielectric
functions.
\begin{acknowledgement}
The author acknowledges Prof. Ding-sheng Wang for his advisements to start the study. The present work has been financially supported by National Science Council, R. O. C. (Project No. NSC92-2811-M-002-041 and NSC93-2811-M-001-065).
\end{acknowledgement}

\begin{table}
\caption{The refractive indices of $\alpha-LiIO_{3}$ crystal at the
static limit. The '$n_{o}$', the '$n_{e}$' and the 'B' label the
ordinary, the extra-ordinary component, and the birefringence,
respectively. The $\lambda$ labels the operating wavelength.}
\label{tab:refra}
\begin{tabular}{llllll}
\hline
Result & $n_{o}$ & $n_{e}$ & B & $\lambda$ (nm)\\
\hline
Present & 2.1127 & 1.7940 & -0.3187 & -\\
Exp1 \cite{Nath} & 1.8385 & 1.7050 & -0.1275 & 2249.3\\
Exp2 \cite{Umegaki} & 1.860 & 1.719 & -0.141 & 1060\\
Exp3 \cite{Takizawa} & 1.7940 & 1.6783 & -0.1157 & 5000\\
\noalign{\smallskip}\hline
\end{tabular}
\end{table}
\end{document}